\documentclass{article}

\usepackage{PRIMEarxiv}

\usepackage[utf8]{inputenc} 
\usepackage[T1]{fontenc}    
\usepackage{hyperref}       
\usepackage{url}            
\usepackage{booktabs}       
\usepackage{amsfonts}       
\usepackage{nicefrac}       
\usepackage{microtype}      
\usepackage{lipsum}
\usepackage{fancyhdr}       
\usepackage{graphicx, subcaption, multirow}       
\graphicspath{{media/}}     

\title{Evaluation of Table Representations to Answer Questions from Tables in Documents : A Case Study using 3GPP Specifications
}

\author{
  Sujoy Roychowdhury, Sumit Soman, HG Ranjani, Avantika Sharma*, Neeraj Gunda*, Sai Krishna Bala \\
  GAIA, Ericsson \\
  Bangalore, Karnataka, India \\
  \texttt{\{sujoy.roychowdhury, sumit.soman, ranjani.h.g, sai.krishna.bala\}@ericsson.com} 
  \thanks{This work was done during the author's internship at GAIA, Ericsson.}
}

\begin{document}
\maketitle

\begin{abstract}
With the ubiquitous use of document corpora for question answering, one important aspect which is especially relevant for technical documents is the ability to extract information from tables which are interspersed with text. The major challenge in this is that unlike free-flow text or isolated set of tables, the representation of a table in terms of what is a relevant chunk is not obvious. We conduct a series of experiments examining various representations of tabular data interspersed with text to understand the relative benefits of different representations. We choose a corpus of $3^{rd}$ Generation Partnership Project (3GPP) documents since they are heavily interspersed with tables. We create expert curated dataset of question answers to evaluate our approach. We conclude that row level representations with corresponding table header information being included in every cell improves the performance of the retrieval, thus leveraging the structural information present in the tabular data. 
\end{abstract}


\section{Introduction}\label{sec:intro}
\begin{figure*}[h!]
    \centering
    \begin{subfigure}[b]{0.93\textwidth}
        \centering
        \fbox{\includegraphics[width=0.93\textwidth]{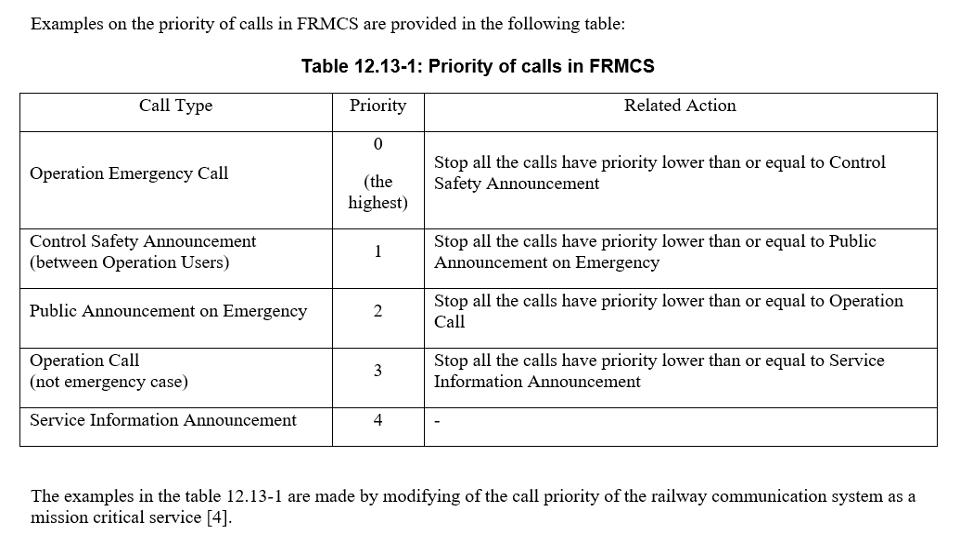}}
        \caption{A snapshot from a 3GPP document showing how table and text are interspersed}
        \label{fig:ex-table}
    \end{subfigure}
    
    \vspace{0.2cm}
    
    \begin{subfigure}[b]{0.93\textwidth}
        \centering
        \fbox{\includegraphics[width=0.93\textwidth]{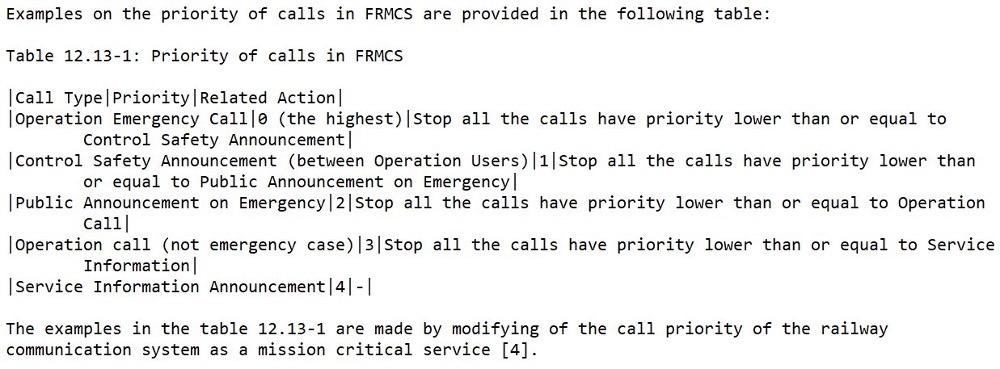}}
        \caption{Table representation in pipe separated format.}
        \label{fig:table_pipe}
    \end{subfigure}
    
    \vspace{0.2cm}
    
    \begin{subfigure}[b]{0.93\textwidth}
        \centering
        \fbox{\includegraphics[width=0.93\textwidth]{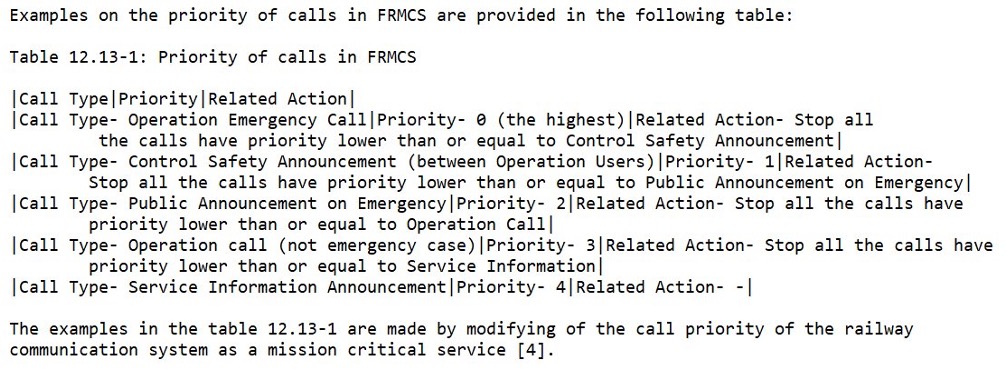}}
        \caption{Table representation in pipe separated format with repeating headers across rows.}
        \label{fig:table_pipe_header}
    \end{subfigure}
    
    \caption{A sample table and its representations used for our experiments.}
    \label{tab:table_representations}
\end{figure*}

Question Answering (QA) based on retrieval from documents has evoked interest from the research community with recent advances in Large Language Models (LLMs) and Retrieval Augmented Generation (RAG) \cite{gao2023retrieval}. Typically questions can potentially be answered from text present in paragraphs or sections of documents. However, questions anchored on technical documents often involve answers that require comprehending information from tables present in these documents. The tables in documents can be of diverse types: numerical entries, textual data, a combination of both, include symbols, abbreviations, technical specifications or terminology, equations and sometimes even images. Some examples of such tables are shown in Figure \ref{fig:ex-table} and Table \ref{tab:3gpp_questions}.

To answer questions related to tables, content from tables in documents must be parsed and represented in a way such that embeddings can be generated. These embeddings in-turn can be used to retrieve most similar chunks to the question embeddings. It is possible that the representations and the choice of embedding models has a bearing on the performance of the QA systems. To the best of our knowledge, a systematic study analysing tabular representations, evaluating on retrieval from technical documents is limited in the literature \cite{ghosh2024robust}. 

The premise of our work is as follows. Prior work has focused on open-domain table QA with training and evaluation on general datasets  \cite{zhao2023localize}. However, domain-specific applications, such as telecom, need technical documents and domain understanding which involve specialized vocabulary. 
$3^{rd}$ Generation Partnership Project (3GPP) documents, \cite{3gpp_release_18} are widely accessed by telecommunications specialists in industry. These include several tables that have technical terminology and related specifications, interspersed with (associated) text, figures and equations. 
It is necessary to extract exact information from the tables to be able to answer questions correctly. 

We would like to highlight few important design choices for our study. 
\begin{itemize}
\item The parsed data considered is textual in nature along with tables (including captions); we do not include images and equations in this work. 
    \item Our study is based on publicly available pre-trained models 
as our study is to create a benchmark of the retrieval performance based on various representations of tabular information.
\item The scope of this study is the retrieval task alone. Current LLMs are able to generate correct answers from contexts containing tables but are dependent on the retrieval output for creating the input context to the LLMs. Hence, it is important to ensure relevant retrievals when the number of tables interspersed with text is large \cite{pan2022end}.

\end{itemize}

We highlight that using publicly available embeddings is not only the starting point of any document retrieval / question-answering task, many industrial applications are often limited to using such embeddings because of cost and data constraints. Our study provides clear recommendations which would be helpful even under such constraints.

\subsection{Research Questions and Contributions}\label{subsec:RQ}

The research questions addressed in this study are as follows:
\begin{itemize}
    \item \textbf{RQ1:} How is the retrieval performance affected when tables are interspersed with text?
    \item \textbf{RQ2:} Does granularity (table or row level) of the embedding chunk impact retrieval performance?
    \item \textbf{RQ3:} Does adding the table header information improve retrieval performance?
    \item \textbf{RQ4:} Would tabular data representation to demarcate columns modify retrieval accuracies?
\end{itemize}
Our contributions from this work are: 
\begin{itemize}
\item We create a QA dataset such that the answers to questions require information from tables (from 3GPP documents).
\item Conduct a systematic study on the effect of table representation on performance on the retrieval task. 
\item We establish that the presence of interspersed text reduces retrieval performance. 
\item We observe that it is better to consider embeddings of rows of tables rather than a single embedding of an entire table. Also, representation introducing table header information along with the tabular content for each cell, prior to the embedding process,  improves performance.
\item We find publicly available pre-trained embedding models to be competitive in terms of retrieval accuracies in an interspersed text and table setting and the table header being introduced in the cell information.  
\end{itemize}

The rest of this paper is organized as follows. 
We position our work vis-\`a-vis that in the literature in Section \ref{sec:litReview}. For a robust evaluation setting, we choose 3GPP's technical specification release 18 with a large number of tables across multiple documents (details in Section \ref{sec:Dataset}). We create an internal dataset curated by Subject Matter Experts (SMEs) that encompass multiple retrieval modalities from the tables through extractive, aggregation and inferential questions. The experimental setup and results are discussed in Sections \ref{sec:Experiments} and \ref{sec:Results} respectively. Section \ref{sec:Conclusions} summarizes results along with possible directions for future work.

\section{Literature Review}\label{sec:litReview}

Methods for representation of tabular data have been prevalent in the literature \cite{badaro2023transformers}. Prior work such as TaPas \cite{herzig2020tapas} and TAPEX \cite{liu2021tapex} retrieve information from a row, given the correct table. In addition, they have been reported to be unsuitable for larger tables due to token length limitations. Representations amenable for RAG tasks have been evaluated to build an end-to-end table QA system \cite{pan2022end}. An approach for converting tables to text and then using it to augment LLMs has been analyzed \cite{min2024exploring}, while knowledge based representations have also been shown to be useful in table QA tasks \cite{hu2024ket, liu2024augment, zhao2023localize}. For handling long representations arising out of complex tables, inner table retrieval has been proposed \cite{lin2023inner}. Identifying relevant parts of tables for QA on large tables has also been evaluated \cite{patnaik2023cabinet}. Our work here complements the literature cited; there are limited works in literature which consider text interspersed with tables, a typical scenario in technical documents. 

\section{Dataset}\label{sec:Dataset}
\begin{figure}[h!]
    \centering
    \resizebox{0.9\columnwidth}{!}{\includegraphics{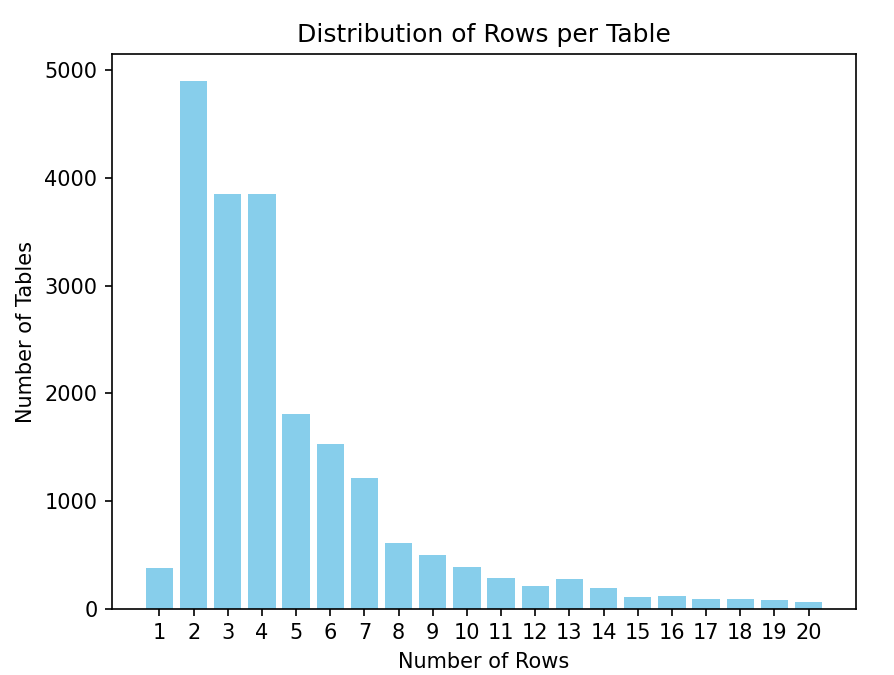}}
    \caption{Histogram for number of rows (excluding header) in table across corpus}
    \label{fig:distribution}
\end{figure}

As mentioned earlier, we use 3GPP Release Specifications as input data; in particular, we consider 3GPP Release 18 \cite{3gpp_release_18}, which contains consider 392 documents. The data is available in \textit{`.docx'} format. Figure \ref{fig:distribution} shows a histogram of the distribution of number of rows of a table (excluding header) - we see that most tables have $<$ 10 rows. This is due to the fact that most tables provide information about certain specifications - the sample image in Figure \ref{fig:ex-table} is such a table.

\begin{table}[h!]
\centering
\resizebox{0.95\columnwidth}{!}{
\begin{tabular}{|l | r|}
\hline
Number of documents                               & 392   \\ 
Number of tables                                  & 21,824 \\ 
Number of sentences in free flowing text            &  948,616     \\ 
Number of tables from which questions are covered & 62    \\ 
Total Number of Questions                         & 278   \\ 
Data Extraction Questions                         & 50    \\ 
Aggregation Questions                             & 77    \\ 
Answers requiring Multiple rows/Columns           & 51    \\ 
Logic based questions                             & 50    \\ \hline
\end{tabular}
}
\caption{Summary statistics of dataset for our experiments.}
\label{tab:datasetStats}
\end{table}

This is a realistic data scenario in contrast to the SciGen dataset \cite{moosavi2021scigen}; SciGen data comprises of scientific tables containing numerical values and their descriptions. The parsed data considered in this work (using 3GPP specifications) includes title, section headings, text, tables, table of contents, comments; header and footer information are ignored. 

For the tabular retrieval evaluation, we have created an internal dataset of 278 questions manually curated by Subject Matter Experts (SMEs). Statistics of our dataset is in Table \ref{tab:datasetStats}. The dataset encompasses multiple retrieval modalities from the tables through extractive, aggregation and inferential questions. A sample example of raw table (Figure \ref{fig:ex-table}) and it's parsed formats are shown in Figs. \ref{fig:table_pipe} and \ref{fig:table_pipe_header} respectively.

We consider four types of questions (the notation used for these types of questions in the following sections of this paper are indicated in parenthesis with the respective question types):
\begin{itemize}
    \item \textbf{Extraction type questions (E)} - requires factual information to be extracted from a single cell in a table.
    \item \textbf{Answers requiring multiple rows and columns (M)} - requires factual information to be extracted, but need access to multiple rows/columns from a single table.
    \item  \textbf{Aggregation based questions (A)} - requires an aggregation of the results across either a row or column of a single table. Aggregation can involve operations such as count or average.
    \item \textbf{Inferential questions (I)} - requires drawing inference from multiple rows and columns of a table. 
\end{itemize}

Sample questions and their corresponding tables and types are provided in Table \ref{tab:3gpp_questions} for reference.

\begin{table*}[t!]
\centering
\begin{tabular}{|p{0.8\textwidth}  p{0.15\textwidth}  p{0.04\textwidth}|}
\hline
\centering \textbf{3GPP Table} & \textbf{Question} & \textbf{Type} \\ \hline
\hfil\multirow{2}{*}[-1.5ex]{\includegraphics[scale=0.30]{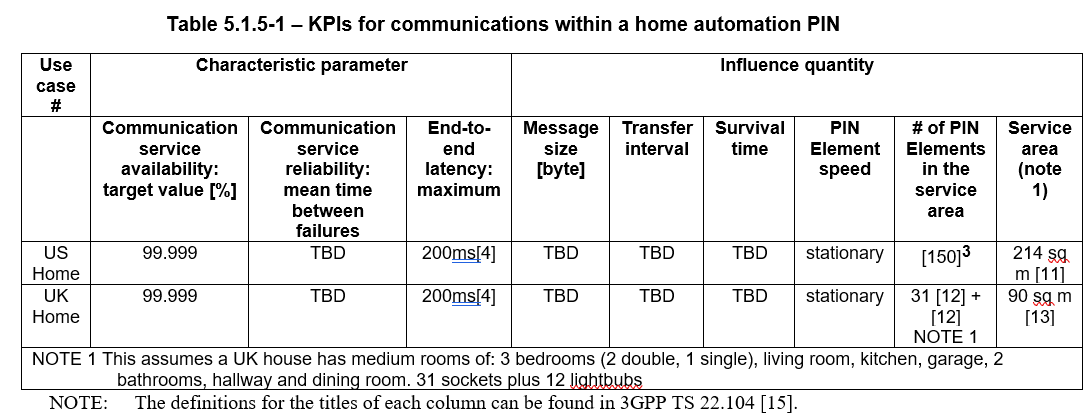}}\hfill & What is the maximum end-to-end latency specified for a UK Home? & E \\ \cline{2-3}
 & How many PIN Elements are there in the service area for the UK Home use case? & E \\ \hline
\hfil\multirow{2}{*}[-1.5ex]{\includegraphics[scale=0.35]{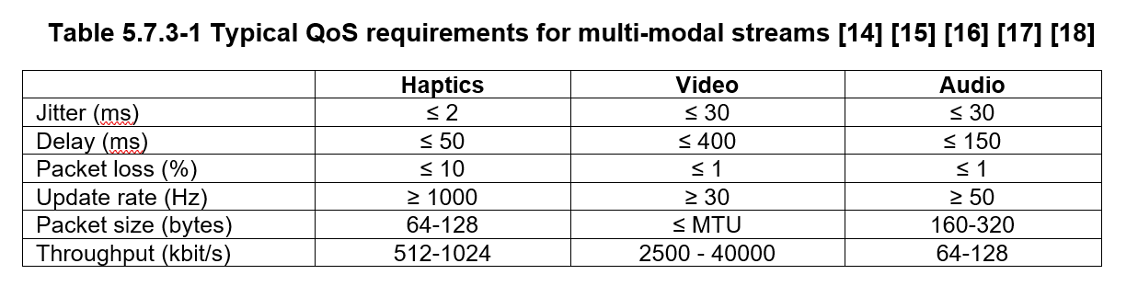}}\hfill & What is the average jitter requirement across all modalities? & A \\ \cline{2-3}
 & Is the throughput requirement for video higher than that for haptics? & I \\ \hline
\hfil\multirow{3}{*}{\includegraphics[scale=0.42]{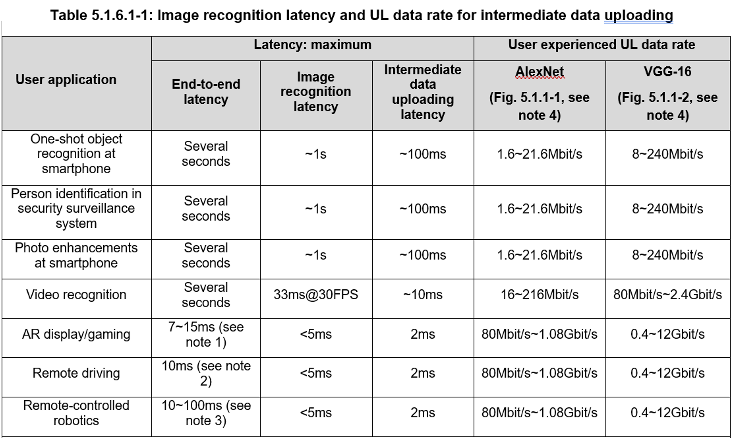}}\hfill & What is the total number of user applications listed in the table? & A \\ \cline{2-3}
 & Identify the user applications with an image recognition latency of less than 5ms. & M \\ \cline{2-3}
 & What is the total range of end-to-end latency for all user applications? & I \\ \hline
\hfil\multirow{2}{*}[-1.5ex]{\includegraphics[scale=0.44]{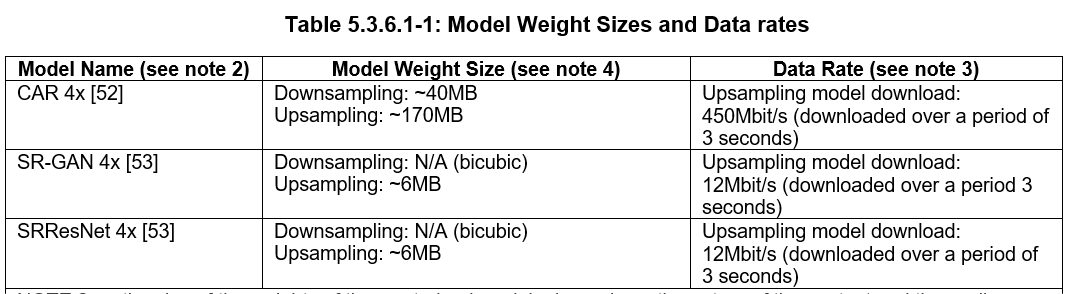}}\hfill & Which models have a bi-cubic downsampling model weight size? & M \\ \cline{2-3}
 & Which model has a lower model weight size for upsampling, CAR 4x or SR-GAN 4x? & I \\ \hline
\end{tabular}
\caption{Sample Questions created from 3GPP Tables (shown in the first column). We also have the type of Question in the third column - A is abbreviation for \textbf{A}ggregation Type of Questions, E for \textbf{E}xtraction, I for \textbf{I}nferential and M for those requiring \textbf{M}ultiple rows and columns}
\label{tab:3gpp_questions}
\end{table*}

\section{Experimental Setup}\label{sec:Experiments}

3GPP \textit{`.docx'} documents are parsed, converted to JSON format using the python-docx library \cite{python_docx}. The JSON representation of a document contains the title, section (and subsection) headings along with free flowing text and tables. The hierarchy and order of the content within the document is preserved in the JSON representation. The parsing process iterates through the elements of the document and identifies sections and tables accordingly. Tables are identified by the \textit{tbl} tag and can be associated to the sections. Table captions are identified by considering the preceding and subsequent elements in order to determine the most likely caption i.e., if the preceding element is a paragraph and starts with \textit{table}, it is most likely the caption for the table. Manual inspection of a large number of tables indicates that most tables have a header row. Based on this observation, the first row of any table has been assumed to contain the header information. We are aware there are small number of tables which contain information like document history, list of associated documents - these have been processed as-is and are part of the corpus for the retrieval task. The header and footer information has been ignored.

\begin{figure*}[h!]
    \centering
    \begin{subfigure}[b]{0.45\textwidth}
        \centering
        \includegraphics[width=\textwidth]{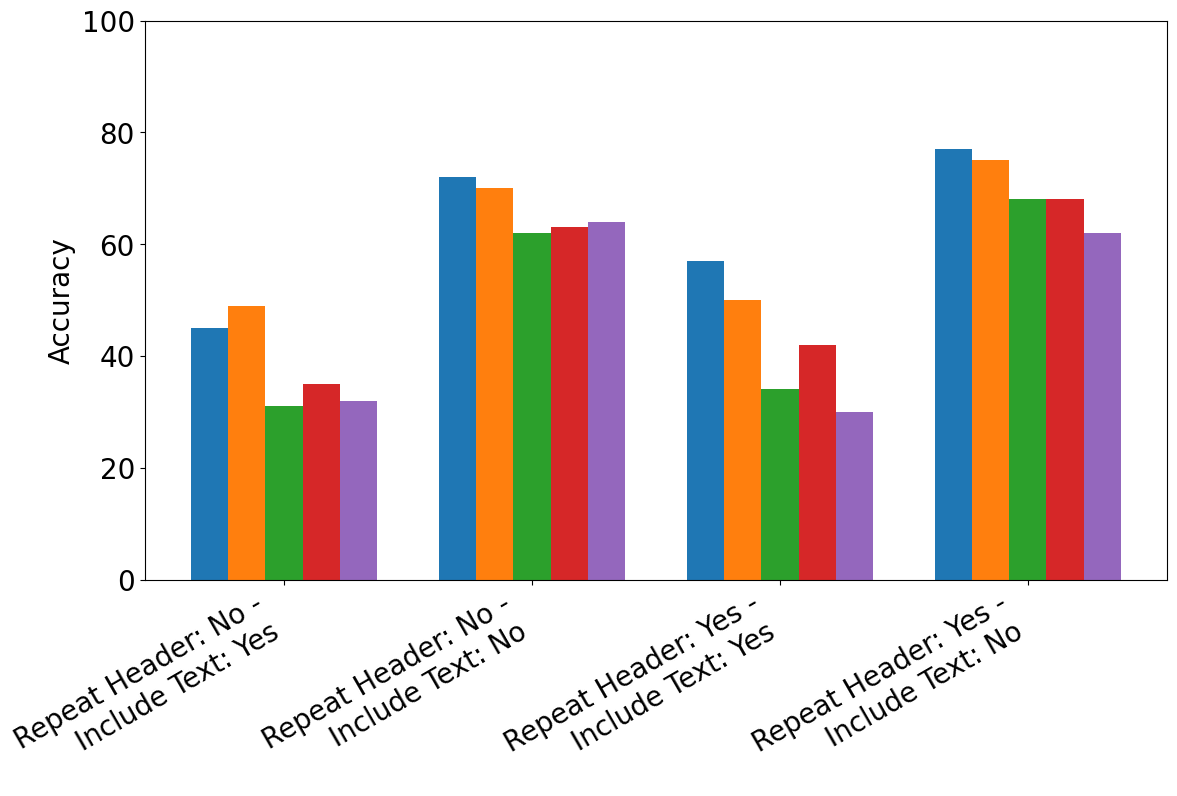}
        \caption{Chunk Level = Table, Separator= Pipe}
        \label{fig:TablePipe}
    \end{subfigure}
    \hfill
    \begin{subfigure}[b]{0.45\textwidth}
        \centering
        \includegraphics[width=\textwidth]{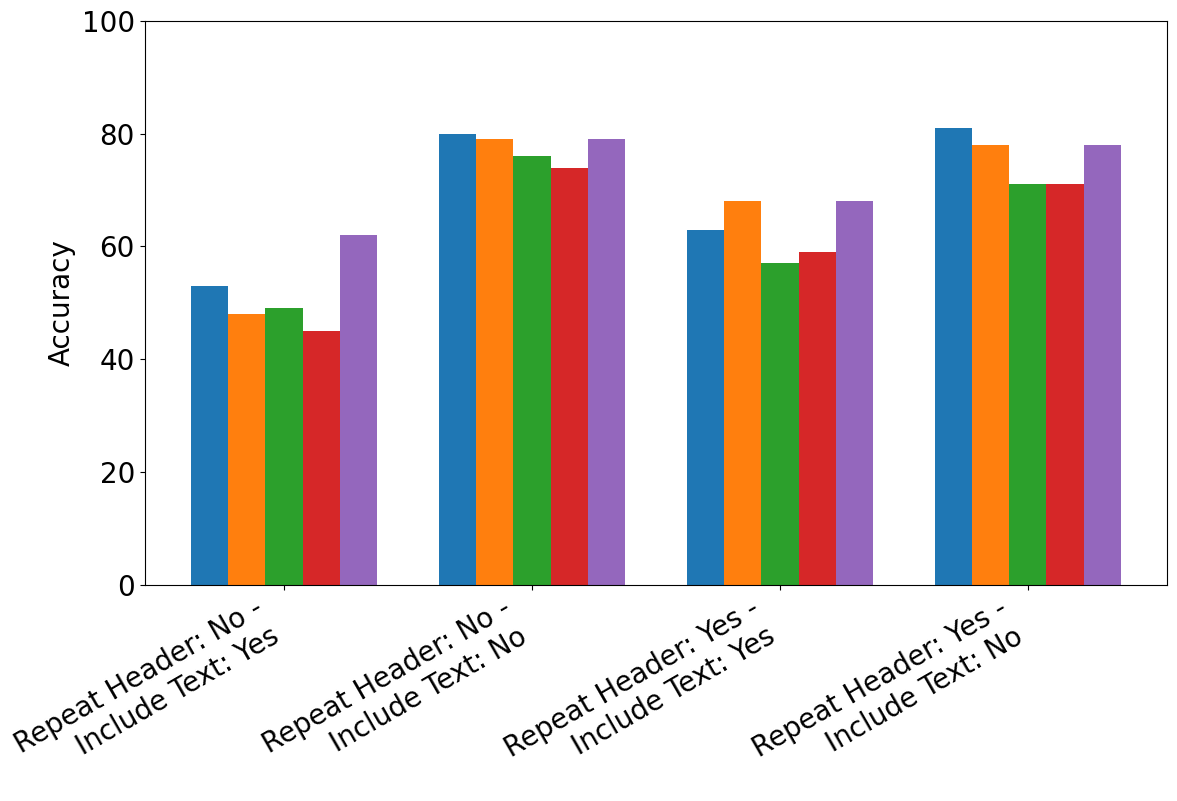}
        \caption{Chunk Level= Row, Separator=Pipe}
        \label{fig:RowPipe}
    \end{subfigure}
    
    \vspace{0.5cm} 

    \begin{subfigure}[b]{0.45\textwidth}
        \centering
        \includegraphics[width=\textwidth]{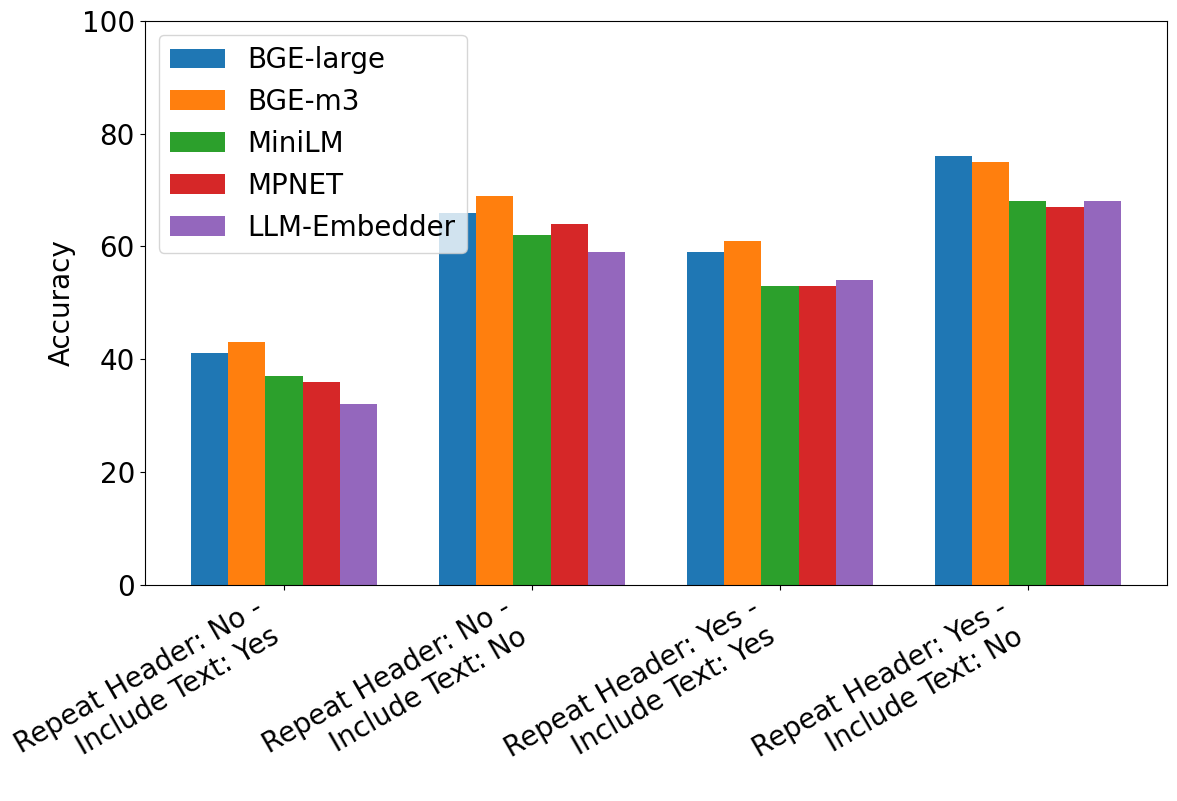}
        \caption{Chunk Level = Table, Separator=Space}
        \label{fig:TableSpace}
    \end{subfigure}
    \hfill
    \begin{subfigure}[b]{0.45\textwidth}
        \centering
        \includegraphics[width=\textwidth]{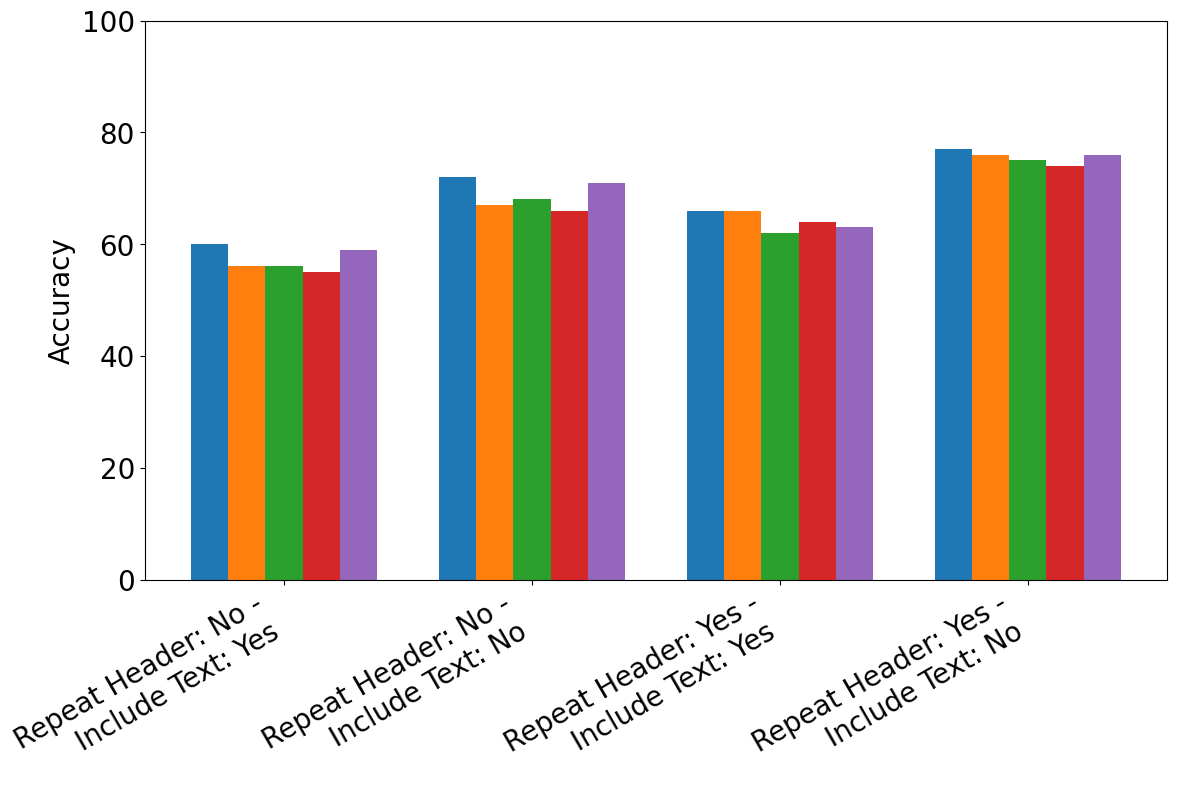}
        \caption{Chunk Level = Row, Separator=Space}
        \label{fig:RowSpace}
    \end{subfigure}
    
    \caption{Comparison of top-5 retrieval accuracy (\%)different representations for tables. The caption given on the bottom left subplot is common for all subplots.}
    \label{fig:all_images}
\end{figure*}
We design our experimental setup to answer the research questions posed in Section \ref{subsec:RQ}. 
First, we study the impact of retrieval based on presence of interspersed text as opposed to using tables alone. 

Second, whilst comparisons of embeddings between sentence and paragraphs has been studied \cite{li2022brief,dong2023open,soman2024observations}, the recommendation for granularity for embeddings of tables is limited in the literature. To this end, we evaluate our embeddings at both a row level chunk and a chunk at the whole table level. 

Third, we consider how best to leverage the structure of tabular information. In this work, we repeat the column heading in every cell - an example is shown in Figure \ref{fig:table_pipe_header}.

Finally, we study if there is an impact of column separators in tabular data representation i.e., does column separators such as space or pipe ($|$) helps. The advantage of pipe separation is that it is less likely to occur within the table - however we are not aware of how it would affect embedding representations using publicly available pre-trained models which may have been less exposed to pipe being present in the data. 

Thus, we have four experimental binary conditions - inclusion (or exclusion) of interspersed text, table vs row level chunking, inclusion (or exclusion) of header in every cell and use of pipe or space as the column separator. We consider a full experimental design of $2^4=16$ parameters and evaluate the retrieval performance for five pre-trained embedding models, \textit{viz.},

\begin{itemize}
    \item From the sentence transformers \cite{reimers2019sentence} library, we consider MPNET \cite{song2020mpnet} and MiniLM (\textit{all-MiniLM-L6-v2}) \cite{wang2020minilm}. {These embeddings have dimensions of 768 and 384 respectively}.
    \item From the BAAI family of embedding models, we consider \textit{bge-large-en} \cite{bge_embedding}, \textit{llm-embedder} \cite{llm_embedder} and \textit{bge-m3} \cite{bge_m3} each of 1024 dimensions
\end{itemize} 

Free flowing text (chunked at a sentence level) interspersed with the tables (chunked at row level or table level) are embedded using the models listed above. We use NLTK's Sentence Tokeniser \cite{loper2002nltk} for creating sentences from free flowing text. We note that all the embeddings have unit $L_2$ norm.

For each of the models considered, we compute the cosine similarity between the question embedding and the sentence embeddings of the corpus formed by the free flowing text and tables' chunks. We retrieve the $k$ chunks (table or sentence) having the highest cosine similarity with the question. If any row, header or caption (corresponding to the correct table) is retrieved in this result, we consider it to be a correct retrieval. Top-$k$ accuracy is the percentage of questions having the correct table association in the $k$ retrieved chunks.

\section{Results}\label{sec:Results}

Figure \ref{fig:all_images} shows the retriever performance for the 16 table representations described in Section \ref{sec:Experiments}. The bars are grouped by the combinations  of inclusion/exclusion of header and text (i.e., Repeat Header: No or Yes, Include text: No or Yes). Figures \ref{fig:TablePipe} and \ref{fig:RowPipe} show the accuracies for pipe-separated representation when the chunking is done at table and row level respectively, while Figures \ref{fig:TableSpace} and \ref{fig:RowSpace} show the accuracies when space is used as a separator for the respective chunking levels.

Comparing retrieval accuracies of `Include Text: No' and `Include Text: Yes' groups across all the 4 sub-plots of Figure \ref{fig:all_images}, we find that retrieval accuracies reduce with introduction of tables interspersed with text. The `Include Text: No' group can be seen as the best possible result (or empirical upper bound) using only tables as input data. This ia the trend seen across models, and across chunking granularities, and data representation. This is the main insight for \textbf{RQ1}. 

Comparing Figures \ref{fig:TablePipe} and \ref{fig:RowPipe}, we find that chunking at row level gives better performance than chunking at table level. This pattern is also observed between Figures \ref{fig:TableSpace} and \ref{fig:RowSpace} and across models. This implies that chunking at row level is preferable for improved retrieval accuracies (\textbf{RQ2}). 

We also notice that when we have tables interspersed with text, accuracies tagged with `Repeat Header: Yes' are higher than `Repeat Header: No'. across most of the models. This establishes that introducing structures in the tables (such as the table header information) can aid in retrieval tasks (\textbf{RQ3}). 

In terms of separator, we find that our results do not indicate consistent change in accuracy, though we obtain the highest performance when pipe is used as a separator (\textbf{RQ4}). 

\begin{figure}[h!]
    \centering
    \resizebox{0.9\columnwidth}{!}{\includegraphics{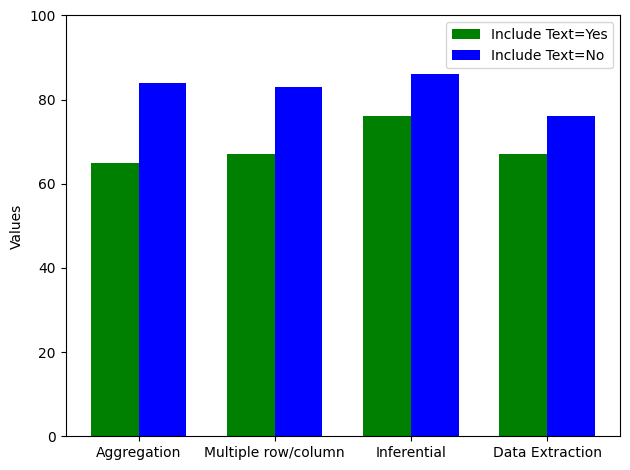}}
    \caption{Performance for the best performing representation (row level chunk, repeated header, pipe separator) with and without interspersed text.}
    \label{fig:distribution_question_type}
\end{figure}

For further insights, we consider only results from row level chunks, and tagged `Include Text: Yes', `Repeat Header: Yes'. This is because these configurations show improvement in retrieval accuracies. 
From Figure  \ref{fig:distribution_question_type}, it can be seen that the drop in performance with inclusion of text is higher for aggregation (A) and multiple row/column retrieval (M) type of questions, highlighting challenges in retrieving relevant portions of text present in multiple chunks. This warrants a more detailed study.


\section{Conclusions}\label{sec:Conclusions}


In this study, we chose a 3GPP corpus having a large number of tables interspersed with text. We curate a dataset of various types of  question-answers from tables and study retrieval performance of various representations across multiple pre-trained embedding models. We observe that the presence of interspersed text creates a reduction in retrieval performance. More importantly, our experiments lead us to conclude that having the corresponding header information to be repeated in every cell of the table prior to embedding it and choosing a separate embedding for every row is beneficial. This exploitation of the structure of tabular data is critical to obtain best performance even on publicly available embedding models. 

Our experiments highlight the importance of our work vis-\`a-vis existing works and datasets that do not consider interspersed text for table retrieval as this is often the case in many practical applications. Although we limit our research to publicly available embeddings, as this is the starting point and often constraints most industrial applications, we expect that the importance of representations would carry through even under domain adaptation techniques.

Our study is limited to a relatively small number of questions because of the fact that a reliable set of useful questions is dependent on SME availability. We wish to expand on our curated dataset over time. This study also can be expanded by following up with domain adaptation techniques - but which may need as a precursor a larger dataset of question answers. Finally, since our approach is generic enough we believe that the observations would hold for other domains too - this is an area of research which can be followed up by us and the community.


\bibliographystyle{unsrt}  
\bibliography{tableQA_arXiv}

\end{document}